\documentclass[conference]{IEEEtran} 
\usepackage{graphicx}
\usepackage{epsfig}
\usepackage{epstopdf}
\usepackage{amsmath,amsthm,amsfonts,amssymb}
\usepackage{multirow}
\usepackage{cite}
\usepackage{verbatim}
\usepackage{tikz}
\usetikzlibrary{arrows}
\usetikzlibrary{backgrounds}
\usepackage{ellipsis}
\usepackage{xcolor}
\usepackage{xifthen}
\usepackage{cancel}
\usepackage{algorithm}
\usepackage{algpseudocode}
\makeatletter
\renewcommand{\fnum@algorithm}{\fname@algorithm}
\makeatother
\theoremstyle{definition} 
\newtheorem{prop}{Proposition}
\theoremstyle{definition} 
\theoremstyle{definition}

\usetikzlibrary{shapes,arrows,intersections}
\usetikzlibrary{patterns,decorations}
\usetikzlibrary{mindmap}
\usetikzlibrary{calc}
\usepackage{ellipsis}
\usetikzlibrary{decorations.pathreplacing,decorations.markings,shapes.geometric,arrows,decorations.pathmorphing,decorations.footprints,fadings,calc,trees,mindmap,shadows,decorations.text,patterns,positioning,shapes,matrix,fit,chains}

\tikzset
{
	BlocksStyle/.style =
	{
		shape			= rectangle,			
		rounded corners	= 0.0cm,				
		minimum height	= 0.7cm,				
		minimum width	= 0.9cm,				
		rotate			= 0,					
		scale			= 1.0,					
		draw			= black,				
		line width		= 0.00cm,				
		align			= center,				
		text			= black,				
		font			= \normalsize\normalfont,	
		inner xsep		= 0.2cm,				
		inner ysep		= 0.2cm,				
	}
}

\tikzset
{
	BlocksStyleb/.style =
	{
		shape			= rectangle,			
		rounded corners	= 0.0cm,				
		minimum height	= 0.7cm,				
		minimum width	= 0.9cm,				
		rotate			= 0,					
		scale			= 1.0,					
		draw			= black,				
		line width		= 0.02cm,				
		align			= center,				
		text			= black,				
		font			= \normalsize\normalfont,	
		inner xsep		= 0.2cm,				
		inner ysep		= 0.2cm,				
	}
}

\tikzset
{
	WideBlocksStyle/.style =
	{
		BlocksStyle,
		text width		= 2.0cm,				
	}
}
\pgfmathsetseed{1}
\usetikzlibrary{shapes,positioning,arrows,calc,graphs}
\usetikzlibrary{decorations.pathreplacing,decorations.markings,shapes.geometric}
\tikzset{naming/.style={align=center,font=\small}}
\tikzset{antenna/.style={insert path={-- coordinate (ant#1) ++(0,0.25) -- +(135:0.25) + (0,0) -- +(45:0.25)}}}
\tikzset{station/.style={naming,draw,shape=dart,shape border rotate=90, minimum width=10mm, minimum height=10mm,outer sep=0pt,inner sep=3pt}}
\tikzset{mobile/.style={naming,draw,shape=rectangle,minimum width=12mm,minimum height=6mm, outer sep=0pt,inner sep=3pt}}
\tikzset{radiation/.style={{decorate,decoration={expanding waves,angle=90,segment length=4pt}}}}

\tikzset
{
	HighlightingStyle/.style =
	{
		color				= black, 
		line width			= 0.04cm,			
		arrows				= -,				
		dotted,
	}
}	

\tikzset
{
	HighlightingStyleD/.style =
	{
		color				= blue,	
		line width			= 0.04cm,			
		arrows				= -,				
		dashed,
	}
}	

\tikzset
{
	HighlightingStyleB/.style =
	{
		color				= black,	
		line width			= 0.04cm,			
		arrows				= -,				
		solid,
	}
}

\tikzset
{
	HighlightingStyleE/.style =
	{
		color				= red,	
		line width			= 0.05cm,			
		arrows				= -,				
		dashed,
	}
}	
\tikzset
{
	HighlightingStyleC/.style =
	{
		color				= black,
		fill=white,
		text=black,	
		line width			= 0.05cm,			
		arrows				= -,				
 		rounded corners		= 0.0cm,			%
		solid,
	}
}

\tikzset
{
	LinesStyle/.style =
	{
		color				= black,	
		line width			= 0.02cm,			
		solid,
	}
}	

\tikzset
{
	LinesStyleC/.style =
	{
		color				= black,	
		line width			= 0.08cm,			
		solid,
	}
}

\tikzset
{
	LinesStyleR/.style =
	{
		color				= black,	
		line width			= 0.08cm,			
		densely dotted,
	}
}

\tikzset
{
	LinesStyleE/.style =
	{
		color				= red,	
		line width			= 0.06cm,			
		solid,
	}
}

\tikzset
{
	LinesStyleb/.style =
	{
		color				= black,	
		line width			= 0.02cm,			
		arrows				= -latex',			
 		shorten <			= 0.1cm,			
		solid,
	}
}

\tikzset
{
	SumNodesStyle/.style =
	{
		shape			= circle,				
		minimum size	= 0.1cm,				
		rotate			= 0,					
		scale			= 0.6,					
		draw			= black,				
		line width		= 0.02cm,				
		text			= black,				
		font			= \normalsize\normalfont,	
	}
}

\DeclareMathOperator{\F2}{\mathbb{F}_2}
\DeclareMathOperator{\sgn}{sgn}
\DeclareMathOperator{\sort}{sort}

\begin{document}

\title{Minimum-Distance Based Construction of Multi-Kernel Polar Codes}

\author{\IEEEauthorblockN{Valerio Bioglio, Fr\'ed\'eric Gabry, Ingmar Land, Jean-Claude Belfiore}
\IEEEauthorblockA{Mathematical and Algorithmic Sciences Lab\\ France Research Center, Huawei Technologies France SASU\\
Email: $\{$valerio.bioglio,frederic.gabry,ingmar.land, jean.claude.belfiore$\}$@huawei.com}} 

\maketitle

\begin{abstract}
In this paper, we propose a construction for multi-kernel  polar codes based on the maximization of the minimum distance. 
Compared to the original construction based on density evolution, our new design shows particular advantages for short code lengths, where the polarization effect has less impact on the performance than the distances of the code.  We introduce and compute the minimum-distance profile and provide a simple greedy algorithm for the code design.  Compared to state-of-the-art punctured or shortened Arikan polar codes, multi-kernel polar codes with our new design show significantly improved error-rate performance.
\end{abstract}


\section{Introduction}
\label{sec:intro}
Polar codes, introduced by Arikan in \cite{polar}, are a new class of channel codes which achieve capacity over various classes of channels under low encoding and decoding complexity.  Also for finite-lengths, these codes show remarkable error rate performance under list decoding.  Only seven years after their discovery, polar codes were recently adopted in the standardization for the control channel of the future 5G system, where the focus is on short-length codes. 

In their original construction, polar codes are based on the polarization effect of the Kronecker powers of the $2 \times 2$ kernel matrix
$T_2 = \tiny\begin{pmatrix}
     1 & 0 \\ 1 & 1 
\end{pmatrix}$.
The generator matrix of a polar code is then a sub-matrix of the transformation matrix $T_2^{\otimes n}$. 
Arikan conjectured in \cite{polar} that the polarization effect is not restricted to powers of the kernel $T_2$, which was verified in \cite{exp_urbanke}, where the authors provide necessary and sufficient conditions for binary kernels $T_p$ of size $p \times p$, $p>2$, to allow for the polarization effect. 
Recently, polar codes based on larger kernels were proposed in \cite{kernel_presman,non_bin_ker}, while in \cite{YB} authors propose to use different kernels of the same size to construct the transformation matrix of the code.

Thanks to these ideas, it is now possible to construct polar codes of any code length of the form $N = p^n$.
However, not all code lengths can be expressed as powers of integers.  To overcome this length matching problem, puncturing \cite{chen_kai_punc}, \cite{isit_punc} and shortening \cite{wang_liu} techniques have been proposed to construct polar codes of arbitrary lengths, at the cost of a loss in terms of polarization speed, and hence worse error rate performance.  

To tackle the code length problem of polar codes, a multi-kernel construction has been proposed in \cite{mk_arxiv}. 
By mixing binary kernels of \emph{different} sizes in the transformation matrix, codes of lengths that are not only powers of integers can be constructed. 
The resulting multi-kernel polar code still benefits from the polarization effect while decoded through successive cancellation \cite{MK_pol_proof}. 
As a result, the new multi-kernel construction largely increases the number of code lengths that can be achieved without puncturing or shortening, with comparable or even better error-rate performance.

For codes based on the polarization effect, the reliability of the input positions is determined by density evolution or other techniques, and then the least reliable positions are frozen. 
This is the design principle of the original construction of polar codes of infinite length \cite{polar}, and it is similarly used for the design of multi-kernel polar codes \cite{mk_arxiv}.  
Such design by reliability is appropriate for long codes under successive cancellation decoding; for short codes under list decoding \cite{list_decoding}, however, design principles that give more weight to distance properties may give superior error-rate performance. 
Related to this is the work in \cite{RM_polar}, where reliability-based design of polar codes for better channels is shown to lead to better distance properties, and ultimately to Reed-Muller codes. 

In this paper, we propose a construction of multi-kernel polar codes that maximizes the minimum distance.    
We show how to find kernels of size larger than 2 that are advantageous in our construction. 
Moreover, we present a simple greedy code design algorithm that maximizes the minimum distance for given kernels. 
Due to the special structure of the kernels of larger size and the resulting flexibility in the code design, our construction of multi-kernel polar codes leads to better distance properties and thus to superior error rate performance under successive cancellation list decoding, compared to the reliability-based construction, and also compared to shortened or punctured codes based on $T_2^{\otimes n}$.

This paper is organized as follows. 
In Section \ref{sec:model}, we review construction, encoding, and decoding of multi-kernel polar codes. 
In Section \ref{sec:design} we describe explicitly the new distance-based design for multi-kernel polar codes. 
In Section \ref{sec:num} we illustrate numerically the performance of the codes, and Section \ref{sec:conclusions} concludes this paper.

\section{Multi-Kernel Polar Codes}
\label{sec:model}
In this section, we briefly review the structure, encoding and decoding of multi-kernel polar codes; 
for details, we refer the reader to \cite{mk_arxiv}. 
Multi-kernel polar codes are a generalization of the Arikan polar codes \cite{polar}, simply referred to as polar codes throughout the paper, and therefore we will provide a comparison to the Arikan construction for clarity.

\subsection{Code Structure and Encoding}

Polar codes are based on the Kronecker product $G_N = T_2^{\otimes n}$, $N=2^n$, where $T_2$ denotes the $2 \times 2$ kernel
\begin{equation*}
  T_2 = \begin{pmatrix} 1 & 0 \\ 1 & 1 \end{pmatrix} .
\end{equation*}
Let us assume an information set $\mathcal{I} \subset [N]$, $[N] = \{0,1,\ldots,N-1\}$, of size $|\mathcal{I}| = K$ and a corresponding frozen set $\mathcal{F} = [N] \backslash \mathcal{I}$ of size $|\mathcal{F}| = N-K$. 
An $(N,K)$ polar code of length $N$ and dimension $K$ is then defined by the encoder $x = u \; G_N$, mapping the input vector $u \in \F2^N$ to the codeword $x \in \F2^N$, where $u_i = 0$ for $i \in \mathcal{F}$, denoting the frozen bits, and $u_i$, $i \in \mathcal{I}$, are the information bits.

Multi-kernel polar codes generalize this construction by mixing binary kernels of different sizes in the Kronecker product forming the transformation matrix. 
Examples of such kernels, which are used in this paper, are
\begin{align}
\label{equ:T3_T5}
  T_3 &= \begin{pmatrix} 1 & 1 & 1 \\ 1 & 0 & 1 \\ 0 & 1 & 1 \end{pmatrix}, &
  T_5 &=    \begin{pmatrix}
              1 & 1 & 1 & 1 & 1 \\ 
              1 & 0 & 0 & 0 & 0 \\
              1 & 0 & 0 & 1 & 0 \\
              1 & 1 & 1 & 0 & 0 \\
              0 & 0 & 1 & 1 & 1
            \end{pmatrix} .  
\end{align}
The transformation matrix of a multi-kernel polar code is given by
\begin{equation}
  G_N = T_{p_1} \otimes T_{p_2} \cdots \otimes T_{p_s} ,
  \label{equ:GN-MK}
\end{equation}
where $T_{p_i}$, $i=1,2,\ldots,s$, denotes the kernel matrix of size $p_i \times p_i$, and kernels of same size can be used multiple times, i.e., it may be that $p_i = p_j$ for some $i,j$. 
The length of the code is $N = p_1 \cdot p_2 \cdots p_s$. 
Note that the ordering of the kernels in the Kronecker product is important for the frozen set design, as the Kronecker product is not commutative. 
An $(N,K)$ multi-kernel polar code is defined by the transformation matrix $G_N$ and the information set $\mathcal{I}$, with corresponding frozen set $\mathcal{F} = [N] \backslash \mathcal{I}$. 
Codewords $x \in \F2^N$ are generated from the input words $u \in \F2^N$ by $x = u \; G_N$, where $u_i = 0$ for $i \in \mathcal{F}$ and $u_i$, $i \in \mathcal{I}$, stores the information bits. 
In \cite{mk_arxiv}, $\mathcal{I}$ is generated according to the reliabilities of the positions in the input vector $u = (u_0,u_1,\ldots,u_{N-1})$, which can be determined e.g. through density evolution \cite{DE_mori}. 
In this case, the information set is composed by the $K$ most reliable positions.  

\begin{figure}[tbh]
\begin{center}
\resizebox{0.51\textwidth}{!}{\begin{tikzpicture}
[
	xscale	= 1,	
	yscale	= 1,	
]

\matrix
(nMatrix)
[
	row sep		= 0.6cm,
	column sep	= 2.1cm, ampersand replacement=\&
]
{

\node (n01) {$u_0$}; \&\node (n02)  {};\& [-4ex]\node (n03) {};\&\node (n04) {};\& [-4ex]\node (n05) {};\&\node (n06) {$x_0$};
	\\
\node (n11) {$u_1$};\&\node(n12) {};\&\node (n13) {};\&\node (n14) {};\&\node (n15) {};\&\node (n16) {$x_1$};\& \\

\node (n21) {$u_2$};\&\node(n22) {};\&\node(n23) {};\&\node (n24) {};\&\node (n25) {};\&\node (n26) {$x_2$};\\
\node (n31) {$u_3$};\&\node(n32) {};\&\node(n33) {};\&\node(n34) {};\&\node (n35) {};\&\node (n36) {$x_3$};\\
\node (n41) {$u_4$};\&\node(n42) {};\&\node(n43) {};\&\node(n44) {};\&\node(n45) {};\&\node (n46) {$x_4$};\&\\
\node (n51) {$u_5$};\&\node (n52) {};\&\node(n53) {};\&\node(n54) {};\&\node(n55) {};\&\node(n56) {$x_5$};\&\\
};


\draw [LinesStyle] (n01) -- (n06) ;

\draw [LinesStyle] (n11) -- ($(n13.west) + (-.3cm,-0cm)$) ;
\draw [LinesStyle] ($(n13.west) + (-.3cm,-0cm)$) --  ($(n23.east) + (.3cm,-0cm)$);
\draw [LinesStyle] ($(n15.west) + (-.3cm,-0cm)$) -- ($(n13.east) + (.3cm,-0cm)$) ;
\draw [LinesStyle] ($(n15.west) + (-.3cm,-0cm)$) -- ($(n35.east) + (.3cm,-0cm)$) ;
\draw [LinesStyle] ($(n15.east) + (.3cm,-0cm)$) -- (n16) ;
\draw [LinesStyle] (n21) -- ($(n23.west) + (-.3cm,-0cm)$) ;
\draw [LinesStyle] ($(n23.west) + (-.3cm,-0cm)$) --  ($(n43.east) + (.3cm,-0cm)$);
\draw [LinesStyle] ($(n25.west) + (-.3cm,-0cm)$) -- ($(n23.east) + (.3cm,-0cm)$) ;
\draw [LinesStyle] ($(n25.west) + (-.3cm,-0cm)$) -- ($(n15.east) + (.3cm,-0cm)$) ;
\draw [LinesStyle] ($(n25.east) + (.3cm,-0cm)$) -- (n26) ;
\draw [LinesStyle] (n31) -- ($(n33.west) + (-.3cm,-0cm)$) ;
\draw [LinesStyle] ($(n33.west) + (-.3cm,-0cm)$) --  ($(n13.east) + (.3cm,-0cm)$);
\draw [LinesStyle] ($(n35.west) + (-.3cm,-0cm)$) -- ($(n33.east) + (.3cm,-0cm)$) ;
\draw [LinesStyle] ($(n35.west) + (-.3cm,-0cm)$) -- ($(n45.east) + (.3cm,-0cm)$) ;
\draw [LinesStyle] ($(n35.east) + (.3cm,-0cm)$) -- (n36) ;
\draw [LinesStyle] (n41) -- ($(n43.west) + (-.3cm,-0cm)$) ;
\draw [LinesStyle] ($(n43.west) + (-.3cm,-0cm)$) --  ($(n33.east) + (.3cm,-0cm)$);
\draw [LinesStyle] ($(n45.west) + (-.3cm,-0cm)$) -- ($(n43.east) + (.3cm,-0cm)$) ;
\draw [LinesStyle] ($(n45.west) + (-.3cm,-0cm)$) -- ($(n25.east) + (.3cm,-0cm)$) ;
\draw [LinesStyle] ($(n45.east) + (.3cm,-0cm)$) -- (n46) ;

\draw [LinesStyle] (n51) -- (n56) ;

\node [coordinate, xshift = -0.3cm, yshift =  0.3cm] (nAux1) at (n03.north west) {};
\node [coordinate, xshift =  0.3cm, yshift =  -0.3cm] (nAux2) at (n53.south east) {};
\draw [HighlightingStyle]  (nAux1) -| (nAux2) -|  (nAux1)
node [below, pos = 0.21] {};

\node [coordinate, xshift = -0.3cm, yshift =  0.3cm] (nAux3) at (n05.north west) {};
\node [coordinate, xshift =  0.3cm, yshift =  -0.3cm] (nAux4) at (n55.south east) {};
\draw [HighlightingStyle] (nAux3) -| (nAux4) -|  (nAux3)
node (Pnode) [below, pos = 0.21] {};

\node [coordinate, xshift = -0.3cm, yshift =  0.2cm] (nAux5) at (n02.north west) {};
\node [coordinate, xshift = 0.2cm, yshift =  -0.2cm] (nAux6) at (n22.south east) {};
\draw [HighlightingStyleB,fill=white] (nAux5) -| (nAux6) -|  (nAux5)
node (Pnode1) [xshift=0.4cm, yshift =-1.4cm] {$T_3$};

\node [coordinate, xshift = -0.3cm, yshift =  0.2cm] (nAux7) at (n32.north west) {};
\node [coordinate, xshift = 0.2cm, yshift =  -0.2cm] (nAux8) at (n52.south east) {};
\draw [HighlightingStyleB,fill=white] (nAux7) -| (nAux8) -|  (nAux7)
node (Pnode1) [xshift=0.4cm, yshift =-1.4cm] {$T_3$};

\node [coordinate, xshift = -0.3cm, yshift =  0.2cm] (nAux9) at (n04.north west) {};
\node [coordinate, xshift = 0.2cm, yshift =  -0.2cm] (nAux10) at (n14.south east) {};
\draw [HighlightingStyleB,fill=white] (nAux9) -| (nAux10) -|  (nAux9)
node (Pnode1) [xshift=0.4cm, yshift =-.9cm] {$T_2$};

\node [coordinate, xshift = -0.3cm, yshift =  0.2cm] (nAux11) at (n24.north west) {};
\node [coordinate, xshift = 0.2cm, yshift =  -0.2cm] (nAux12) at (n34.south east) {};
\draw [HighlightingStyleB,fill=white] (nAux11) -| (nAux12) -|  (nAux11)
node (Pnode1) [xshift=0.4cm, yshift =-.9cm] {$T_2$};

\node [coordinate, xshift = -0.3cm, yshift =  0.2cm] (nAux13) at (n44.north west) {};
\node [coordinate, xshift = 0.2cm, yshift =  -0.2cm] (nAux14) at (n54.south east) {};
\draw [HighlightingStyleB,fill=white] (nAux13) -| (nAux14) -|  (nAux13)
node (Pnode1) [xshift=0.4cm, yshift =-0.9cm] {$T_2$};

%
%

\node[left of = Pnode,node distance = 1.7cm] {Stage 1};
\node[left of = Pnode,node distance = 5.8cm] {Stage 2};
\end{tikzpicture}}
\caption{Tanner graph of the multi-kernel polar code for $G_6 = T_2 \otimes T_3$.}
\label{fig:G_6}
\end{center}
\end{figure}
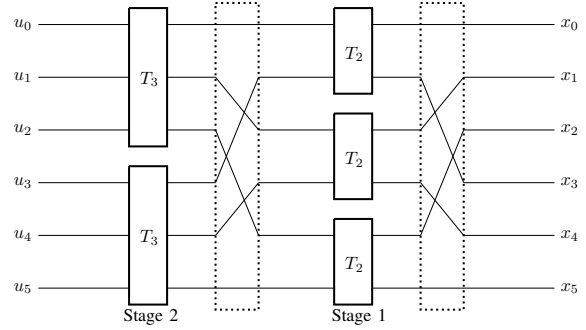

Similar to polar codes, the Tanner graph of multi-kernel polar codes can be constructed. 
While the Tanner graph of polar codes consists solely of $2 \times 2$ blocks, each corresponding to the kernel $T_2$, the Tanner graph of multi-kernel polar codes consists of various blocks, corresponding to the different kernels used. 
The Tanner graph for the transformation matrix in \eqref{equ:GN-MK} consists of $s$ stages. 
On stage $i$, there are $N/p_i$ blocks, each of size $p_i \times p_i$, corresponding to a $T_{p_i}$ kernel, with $p_i$ edges to the left and to the right. 
The connections and edge permutations follow from the Kronecker product \cite{mk_arxiv}. 
An example is given in Fig.~\ref{fig:G_6}, with edge-permutations indicated by dotted boxes.

\subsection{Decoding of Multi-Kernel Polar Codes}

Decoding of multi-kernel polar codes is performed similarly to polar codes, using successive cancellation (SC) decoding on the Tanner graph of the code \cite{mk_arxiv}, or enhanced SC-based decoding methods like SC list (SCL) decoding \cite{list_decoding}.  
Log-likelihood ratios (LLRs) are passed along the Tanner graph from the right to the left, while hard decisions on decoded bits are passed from the left to the right. 
The major difference to decoding of polar codes is given by the computations in the blocks corresponding to the new kernels. 

\begin{figure}[bht]
  \begin{center}
	\begin{tikzpicture}[scale=0.2, line width=0.8pt]
	  \draw   (0,0) rectangle (4,8) node[midway]{$T_p$};
	  \draw[] (-2,7) node[left] {$u_0 , \lambda_0$} -- (0,7) ;
	  \draw[] (-2,5) node[left] {$u_1 , \lambda_1$} -- (0,5) ;
	  \draw[] (-0.5,3.5) node[left] {$\vdots$}    (0,3) ;
	  \draw[] (-2,1) node[left] {$u_{p-1} , \lambda_{p-1}$} -- (0,1) ;
	  \draw[] (4,7)                     -- (6,7) node[right] {$x_0 , L_0$} ;
	  \draw[] (4,5)                     -- (6,5) node[right] {$x_1 , L_1$} ;
	  \draw[] (4,3)                        (4.5,3.5) node[right] {$\vdots$} ;
	  \draw[] (4,1)                     -- (6,1) node[right] {$x_{p-1} , L_{p-1}$} ;
	\end{tikzpicture}
  \end{center}
  \caption{Block in Tanner graph, corresponding to $p \times p$ kernel $T_p$.}
  \label{fig:block-Tl}
\end{figure}
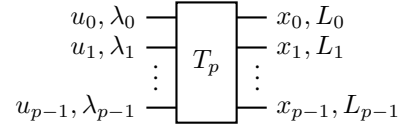

The notation used for the $p \times p$ block corresponding to a $T_p$ kernel is depicted in Fig.~\ref{fig:block-Tl}. 
Denote $u = (u_0,u_1,\ldots,u_{p-1})$ the binary input vector to this block and $x = (x_0,x_1,\ldots,x_{p-1})$ its binary output vector. 
Then we have the relationship $u \; T_p = x$, defining the update rule for the hard-decisions going from left to right. 
Further denote $L_i$ the LLR of output bit $x_i$ and $\lambda_i$ the LLR of the input bit $u_i$.  
The general structure of the update rule for LLRs, going from right to left, is  $ \lambda_i = f( L_0 , L_1 , \ldots , L_{l-1} , \hat{u}_0 , \hat{u}_1 , \ldots, \hat{u}_{i-1} )$, i.e., all LLRs $L_j$ and only previous hard-decisions (estimates) $\hat{u}_j$ may be used for the computation, following the SC principle. 
The corresponding LLR calculations for $T_2$ from \cite{polar} are
\begin{align*}
  \lambda_0 &=  L_0 \boxplus L_1  , \\
  \lambda_1 &= (-1)^{u_0} \cdot L_0 + L_1,
\end{align*}
for $T_3$ depicted in (\ref{equ:T3_T5}), from \cite{mk_arxiv}, are
\begin{align*}
  \lambda_0 &=  L_0 \boxplus L_1 \boxplus L_2  , \\
  \lambda_1 &=  (-1)^{u_0} \cdot L_0 + L_1 \boxplus L_2  , \\
  \lambda_2 &=  (-1)^{u_0} \cdot L_1 + (-1)^{u_0 \oplus u_1} \cdot L_2 ,
\end{align*}
and for $T_5$ in (\ref{equ:T3_T5}), presented here for the first time, are
\begin{align*}
  \lambda_0 &= L_1 \boxplus L_2 \boxplus L_4  , \\[1ex]
  \lambda_1 &= (-1)^{\hat{u}_0} \cdot ( L_0 \boxplus (L_2 + (L_1 \boxplus L_4)) \boxplus L_3 )  , \\[1ex]
  \lambda_2 &= (-1)^{\hat{u}_1 } \cdot ( L_0 \boxplus L_1 )  + ( L_3 \boxplus L_4 )  , \\[1ex]
  \lambda_3 &= (-1)^{\hat{u}_0 \oplus \hat{u}_1 \oplus \hat{u}_2} \cdot L_0  + (-1)^{\hat{u}_0} \cdot L_1  + ( L_2 \boxplus (L_3 + L_4) )  , \\[1ex]
  \lambda_4 &= (-1)^{\hat{u}_0 \oplus \hat{u}_3} \cdot L_2   + (-1)^{\hat{u}_0 \oplus \hat{u}_2} \cdot L_3   + (-1)^{\hat{u}_0} \cdot L_4 .
\end{align*} 
The boxplus operator for two LLRs $a$ and $b$ may be evaluated exactly as $ a \boxplus b = 2 \tanh^{-1} \bigl( \tanh\frac{a}{2} \cdot \tanh\frac{b}{2} \bigr)$ or approximately as $ a \boxplus b \approx \sgn a \cdot \sgn b \cdot \min\{|a|,|b|\}$; the extension to multiple LLRs is as usual.  For other kernels similar LLR update equations can be derived \cite{arb_ker}. 
In the next section, we will show that the presented kernels $T_3$ and $T_5$ permit to construct multi-kernel polar codes with good minimum distance.

\section{Design for Minimum-Distance}
\label{sec:design}
In this section, we describe how to design multi-kernel polar codes to increase the minimum distance of the code.
In \cite{mk_arxiv}, the information set $\mathcal{I}$ is selected according to reliability. 
This approach, which is also commonly followed for polar codes, is optimal for SC decoding when the code length tends to infinity. 
For short codes, however, the distance properties are more crucial than the polarization effect.

In the following, we focus on multi-kernel polar codes with transformation matrix of the form $G_N = T_2^{\otimes n} \otimes T_p$, i.e., a polar code composed with a larger kernel at the end.  
This is not a very limiting assumption for the following reasons.
First, the proposed design is to be used for short codes, for which the use of a single larger kernel is usually sufficient. 
Second, the larger kernel can be the composition of multiple smaller kernels. 
And third, changing the order of the kernels in the Kronecker product is equivalent to a row and column permutation of $G_N$, and thus leads to equivalent codes.

\subsection{Minimum-Distance Spectrum}

In the following, we will determine the minimum distance $d$ achievable by a code generated by  selecting $K$ rows of a transformation matrix $G_N$. 
More formally, we define the \emph{minimum-distance spectrum} $S_{G_N}$ of the transformation matrix $G_N$ to be the mapping from dimension $K$ to the maximal minimum distance $d$ achievable by selecting an information set $\mathcal{I}$ of size $K$, i.e., $S_{G_N}(K)$ is the largest minimum distance achievable by an $(N,K)$ multi-kernel polar code derived from the transformation matrix $G_N$.

Finding the minimum-distance spectrum of a code is in general a complex task, which may be accomplished e.g. by an exhaustive search. 
Under certain constraints, however, the minimum-distance spectrum of a multi-kernel polar code can be easily calculated based on the minimum-distance spectra of its building kernels.  
In fact, for polar codes, 
\begin{equation*}
  S_{T_2^{\otimes n}} = \sort([2 \quad 1]^{\otimes n}) , 
\end{equation*}
where $\sort(x)$ is the vector $x$ sorted in decreasing order, since polar codes have the same transformation matrix as Reed-Muller codes. 
In the following, we prove that a similar property holds for multi-kernel polar codes, allowing one to calculate the minimum-distance spectrum of the transformation matrix $G_N$ using the Kronecker product of the spectra of the kernels composing it. 

\begin{prop}[Minimum-distance spectrum] 
\label{prop:spectrum} \rule{1ex}{0ex}\\
If $G_N = T_2^{\otimes n} \otimes T_p$, then $S_{G_N} = \sort(S_{T_2^{\otimes n}} \otimes S_{T_p})$.

\begin{proof}
The proposition is proved by induction on the number $n$ of $T_2$ kernels employed in the transformation matrix $G_N$. 
The property obviously holds for $n=0$, and by inductive hypothesis we suppose that $S_{G_{N/2}} = \sort(S_{T_2^{\otimes n-1}} \otimes S_{T_p})$ given $G_{N/2} = T_2^{\otimes n-1} \otimes T_p$.
Given the transformation matrix $G_N = T_2^{\otimes n} \otimes T_p = \tiny\begin{pmatrix} G_{N/2} & 0 \\ G_{N/2} & G_{N/2} \end{pmatrix}$, this matrix can be divided into two parts, an upper matrix $G^U = [G_{N/2} | \mathbf{0}]$ and a lower matrix $G_L = [G_{N/2} |G_{N/2} ]$, for which $S_{G^U} = S_{G_{N/2}}$ and $S_{G^L} = 2 S_{G_{N/2}}$. 
Given $V = \sort(S_{T_2^{\otimes n}} \otimes S_{T_p})$, the goal of the proof is to show that $S_{G_N} = V$, i.e., that for every dimension $K$, there exists a subset of $K$ rows of $G_N$ such that the span of these rows has minimum distance $V(K)$.

To do that, for every $K$ we show how to construct a sub-matrix of $G_N$ for which all the vectors of its span have Hamming weight not smaller than $V(K)$. 
In fact, by construction, for every $K$ there exist two integers $K^U$ and $K^L$ such that $K^U + K^L = K$, and two sub-matrices $G_A^U$ and $G_B^L$, formed by $K^U$ rows of $G^U$ and by $K^L$ rows of $G^L$ respectively, such that $S_{G_A^U}(K^U) \geq V(K)$ and $S_{G_B^L}(K^L) \geq V(K)$. 
To end the proof, it is sufficient to use the distance property of the classical $(u | u+v)$ construction \cite{McW-Sloane} to verify that the code generated by $G_{A,B} = \left[ \frac{G_A^U}{G_B^L} \right]$ has minimum distance $\min(S_{G_A^U}(K),S_{G_B^L}(K)) = V(K)$. 
\end{proof} 
\end{prop}

The proposition shows how to exploit the spectra of the building kernels to evaluate the minimum-distance spectrum of the multi-kernel polar code. 
Moreover, the constructive nature of the proof suggests a greedy technique to build multi-kernel polar codes with optimal minimum distance. 
Before describing the algorithm in detail, in the following section we present kernel design principles leading to codes with good minimum distance spectra. 

\subsection{Kernel Design}

For polar codes, kernels are usually designed to maximize the polarization effect on the input bits of the transformation $G_N$, and the information positions are then selected in reliability order. 
For short codes, however, the polarization effect is far less important than distances of the code, and kernels should be designed taking this aspect into account. 
Different kernels have different spectra, while polar codes are limited by the spectrum of the kernel $T_2$. 
Multi-kernel polar codes permit to create codes of desired minimum distance by changing the kernels composing the transformation matrix. 
If the kernels are designed properly, the information set can then be selected such that a large minimum distance is achieved for the desired length and dimension. 

As an example, consider the $T_3$ kernel depicted in (\ref{equ:T3_T5}), introduced in \cite{mk_arxiv}, and its minimum-distance spectrum.
For the information set of size 1, one row has to be selected: in order to maximize the minimum distance, the first row, $(1 \; 1 \; 1)$, is selected, giving minimum distance 3; any other row selection would result in a smaller minimum distance, namely 2. 
For the information set of size 2, the last two rows, $(1 \; 0 \; 1)$ and $(0 \; 1 \; 1)$, are selected, generating a code of minimum distance 2; any other row selection would result in a smaller minimum distance. 
Finally, for a code of dimension 3, all rows have to be selected, resulting in a code of minimum distance 1. 
$T_3$ thus has the minimum-distance spectrum $S_{T_3}=(3,2,1)$. 
As opposed to that, the construction by reliability selects the last row for dimension 1, the last two rows for dimension 2, and all rows for dimension 3; this gives minimum-distance spectrum $(2,2,1)$.  
The proposed $T_5$ kernel presents a similar behavior, with minimum-distance spectrum $S_{T_5}=(5,3,2,1,1)$. 

\subsection{Greedy Row-Selection Algorithm}

In the previous sections, we described how to calculate the minimum-distance spectrum of the transformation matrix of a certain class of multi-kernel polar codes. 
The scope of this section is to describe how to determine the actual information set that achieves this minimum distance. 
As for the minimum-distance spectrum itself, this may be accomplished by an exhaustive search, which in general will be very complex. 
The proof of Proposition~\ref{prop:spectrum}, however, gives an insight on how to select rows of $G_N$ to achieve the minimum-distance spectrum.  

In the following we describe a greedy algorithm able to accomplish this task; the pseudo code is provided in Algorithm~\ref{algo}. 
Since the algorithm is based on Proposition~\ref{prop:spectrum}, it finds an optimal solution if only one kernel of size larger than 2 is used in the construction, and this kernel is the last term in the Kronecker product.
The algorithm may as well be applied in the case of multiple kernels of size larger than 2, also at the end of the Kronecker product, by treating the Kronecker product of these kernels as one large kernel, for which the minimum-distance spectrum has to be determined before the algorithm is applied. 

Given a transformation matrix $G_N = T_2^{\otimes n} \otimes T_p$, we assume the kernel $T_p$ to have minimum-distance spectrum $S_{T_p}= (d_p (1), \cdots, d_p (p))$, where $d_p (k)$ is the minimum distance of the code of dimension $k$. 
The list $I^k = \{i^k_1,\dots,i^k_k\}$ is associated to every entry $d_p (k)$ of the spectrum, collecting the indices of the $k$ rows of $T_p$ giving the optimal minimum distance of the kernel. 
To begin with, the vector $r_N = (2,1)^{\otimes n} \otimes S_{T_p}$ is created. 
This vector is an unsorted version of the minimum-distance spectrum, collecting the minimum achievable distances of each part of $G_N$. 

For a code of dimension $K$, at each step the algorithm adds sequentially one row index to the information set $\mathcal{I}$, which is initially empty. 
At each step, the position $l$, with $l = 0,\dots,N-1$, of the last largest entry in $r_N$ is found, and $r_N (l)$ is set to zero.  
After that, the value $c = (l \: \text{mod} \: p)+1$ and $q = l-c+1$ are calculated, giving the row position within the kernel and the row index in the transformation matrix where the corresponding kernel starts, respectively. 
In fact, since in $S_{T_p}$ the distances are sorted in descending order, we know that $\{i^{c}_1+q,\dots,i^{c}_{c}+q\} \subset \mathcal{I}$. 
The algorithm deletes these $c$ indices belonging to $I^{c}$, substituting them with the $c+1$ indices given by $I^{c+1}$; by the constructive proof of Proposition~\ref{prop:spectrum}, we know that the resulting code has the desired minimum distance. 
Of course, if $c=0$, no rows of that part of the matrix are already in the information set, and therefore no information indices are deleted. 
In practice, at each step the information set is updated as $\mathcal{I} = \mathcal{I} \setminus \{i^{c}_1+q,\dots,i^{c}_{c}+q\} \cup \{i^{c+1}_1+q,\dots,i^{c+1}_{c+1}+q\}$. 
The algorithm stops when $\mathcal{I}$ includes $K$ elements. 
The remaining $N-K$ indices compose the frozen set $\mathcal{F}$. 
\begin{algorithm}[htb]
\caption{Information set to maximize minimum distance} \label{algo}
\begin{algorithmic}[1]
\State $\text{Initialize the set } \mathcal{I} = 0$
\State $\text{Load } N \text{-vector } r_N$
\State $\text{Load } j \text{-vectors } I^j, \quad j = 1 \dots p$
\For{$k = 1 \dots K$}
   \State $l = \text{argmax} (r_N)$
   \State $c = (l \: \text{mod} \: p)+1$
   \State $q = l-c+1$
   \State $r_N(l) = 0$
   \If{$c>1$}
      \State $\mathcal{I} = \mathcal{I} \setminus \{i^{c}_1+q,\dots,i^{c}_{c}+q\} \cup \{i^{c+1}_1+q,\dots,i^{c+1}_{c+1}+q\}$
   \Else
      \State $\mathcal{I} = \mathcal{I} \cup \{i^{1}_1\}$
   \EndIf
\EndFor
\end{algorithmic}
\end{algorithm}

\subsection{Construction Example}
To illustrate our construction, in the following we describe the minimum distance design of a multi-kernel polar code of length $N = 6$ depicted in Fig.~\ref{fig:G_6} with transformation matrix 
  \begin{displaymath}
    G_6 = T_2 \otimes T_3 
      = \begin{pmatrix}
          T_3 & 0 \\ T_3 & T_3    
        \end{pmatrix}
        = \begin{pmatrix} 1 & 1 & 1 & 0 & 0 & 0 \\ 
                          1 & 0 & 1 & 0 & 0 & 0\\
                          0 & 1 & 1 & 0 & 0 & 0\\
                          1 & 1 & 1 & 1 & 1 & 1\\
                          1 & 0 & 1 & 1 & 0& 1 \\
                          0 & 1 & 1 & 0 & 1& 1 \\  
          \end{pmatrix}  .
  \end{displaymath}
For the described kernel of size 3, we have that $S_{T_3} = (3,2,1)$ with $\mathcal{I}_{T_3}(1) = \{0\}$, $\mathcal{I}_{T_3}(2) = \{1,2\}$ and obviously $\mathcal{I}_{T_3}(3) = \{0,1,2\}$. 
The minimum-distance spectrum is given by $S_{G_6} = \sort((2,1) \otimes S_{T_3}) = (6,4,3,2,2,1)$; consequently, $r_6 = (3,2,1,6,4,2)$. 
It is worth noticing that the minimum-distance spectrum of the reliability construction is $(4,4,2,2,2,1)$. 
If a rate 1/2 code has to be designed, the positions of the $K=3$ information bits are needed. 
The information set $\mathcal{I}$ is initially empty. 
At the first step, $l=3$, hence $c=0$ and $q=3$; since $c=0$, no entries of $\mathcal{I}$ have to be deleted, and $\mathcal{I} = \{3\}$. 
At the second step, $l=4$, so $c=1$ and $q=3$; the information set is calculated as $\mathcal{I} = \mathcal{I} \setminus \{3\} \cup \{4,5\} = \{4,5\}$. 
Finally, at the third step $l=0$, and the resulting information set is $\mathcal{I} = \{4,5\} \cup \{0\} = \{0,4,5\}$. 
A comparison of the information sets calculated by the proposed algorithm following the distance criterion and the one resulting from the reliability order is presented in Table~\ref{table:comparison} for various dimensions $K$. 
We observe that the proposed design always outperforms the reliability-based designs in terms of minimum distance, or performs identically when the reliability-based construction is equivalent to the minimum-distance based construction.

\begin{table}[h]
\begin{center}
\resizebox{8.5cm}{!}{
  \begin{tabular}{| c | c | c | c | c |}
  \hline
  & Rate & $1/6$ & $2/6$ & $3/6$ \\ \hline
  Reliability & Information Set & $u_5$ & $(u_4,u_5)$ &  $(u_2,u_4,u_5)$ \\
  Design & Minimum Distance & 4 & 4 & 2   \\ \hline
  Distance & Information Set & $u_3$ & $(u_4,u_5)$ & $(u_0,u_4,u_5)$ \\ 
  Design & Minimum Distance & \bf{6} & 4 & \bf{3} \\ \hline
\end{tabular}}
\end{center}
\caption[]{Comparison of minimum distances for $N=6$.}
\label{table:comparison}
\end{table}

\section{Numerical Illustrations}
\label{sec:num}
\begin{figure}[tb]
\centering
\includegraphics[width=0.51\textwidth]{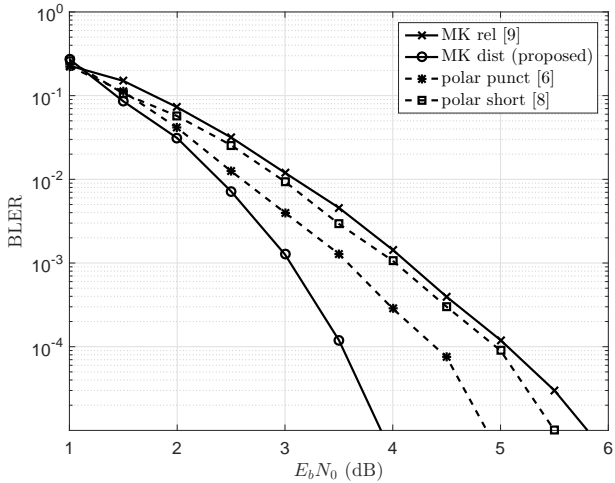}
\vspace*{-0.5ex}
\caption{Block error rates for length $N = 192$ and rate $R = 1/2$ under SCL decoding with list size $L=8$.}
\label{fig:plot_192}
\end{figure}

\begin{figure}[tb]
\centering
\includegraphics[width=0.51\textwidth]{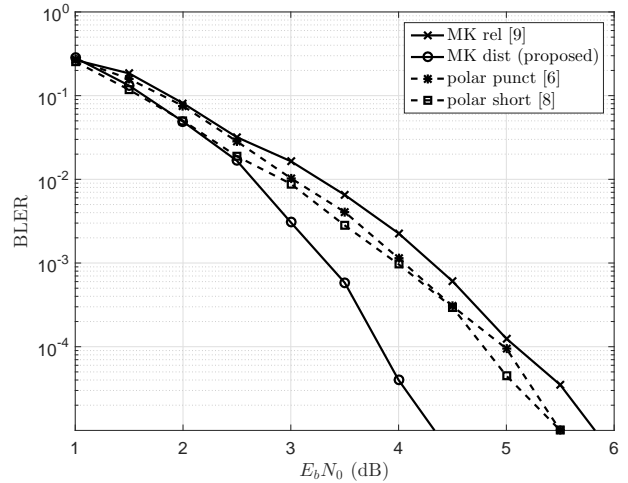}
\vspace*{-0.5ex}
\caption{Block error rates for length $N = 144$ and rate $R = 1/2$ under SCL decoding with list size $L=8$.}
\label{fig:plot_144}
\end{figure}

\begin{figure}[tb]
\centering
\includegraphics[width=0.51\textwidth]{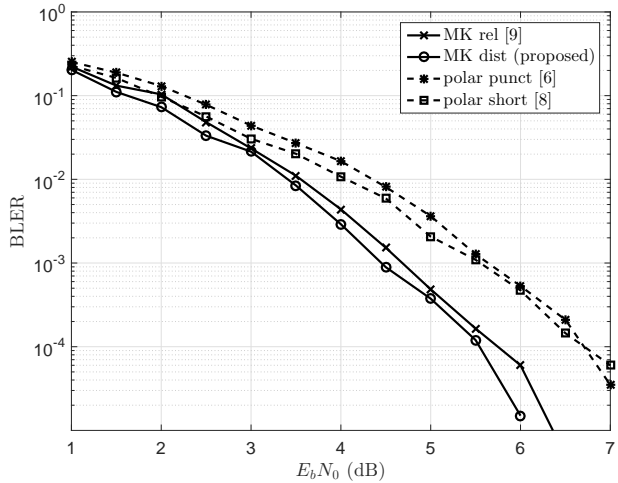}
\vspace*{-0.5ex}
\caption{Block error rates for length $N = 40$ and rate $R = 1/2$ under SCL decoding with list size $L=8$.}
\label{fig:plot_40}
\end{figure}

\begin{figure}[tb]
\centering
\includegraphics[width=0.51\textwidth]{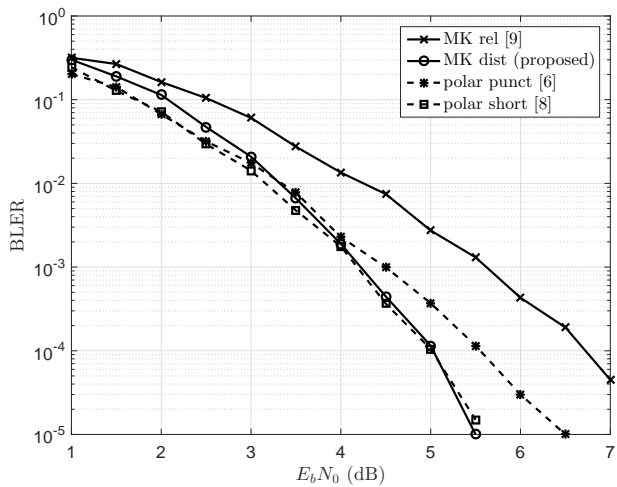}
\vspace*{-0.5ex}
\caption{Block error rates for length $N = 90$ and rate $R = 1/2$ under SCL decoding with list size $L=8$.}
\label{fig:plot_90}
\end{figure}

In the following, we show the performance of the proposed minimum distance construction of multi-kernel polar codes. 
In particular, in Figures \ref{fig:plot_192}, \ref{fig:plot_144}, \ref{fig:plot_40} and \ref{fig:plot_90}, we show the BLER performance of the codes designed according to the proposed minimum distance  construction under list decoding \cite{list_decoding} with list size $L=8$ for BPSK transmission over an additive white Gaussian noise (AWGN) channel. 
Our proposal, coined \emph{MK-dist} in the figures, will be compared to the reliability-based design of multi-kernel polar codes proposed in \cite{mk_arxiv}, coined \emph{MK-rel} in the figures. 
We emphasize that MK-dist is designed according to the row selection algorithm described in the previous sections. 
Moreover, we add as references state-of-the-art punctured \cite{chen_kai_punc} and shortened \cite{wang_liu} polar code constructions, coined \emph{polar-punct} and \emph{polar-short} respectively. 

First, in Figure \ref{fig:plot_192}, we show the performance of a code with length $N = 192$ and dimension $K = 96$. 
In this case, the transformation matrix is given by $T_{192}=T_2^{\otimes 6} \otimes T_3$, i.e., there is only one $T_3$ kernel at the rightmost of the Kronecker product. 
In this case, the proposed row selection algorithm can be run using the minimum-distance spectrum of the $T_3$ kernel presented before, and we observe that the minimum-distance based design outperforms all other designs. 

In Figure \ref{fig:plot_144} we show the performance of a code of length $N = 144$ and dimension $K = 72$. 
The transformation matrix is given by $T_{144}=T_2^{\otimes 4} \otimes T_3^{\otimes 2}$, i.e., there are two $T_3$ kernels at the rightmost of the Kronecker product. 
In this case, the minimum-distance spectrum for the Kronecker product kernel $T_3 \otimes T_3$ has to be calculated, along with the auxiliary lists $\mathcal{I}_{T_3 \otimes T_3}$. 
Potentially, two different kernels of size 3 may be used, like proposed in \cite{YB}, augmenting the flexibility of the minimum-distance spectrum, but this kind of optimization is out of the scope of this paper. 
The resulting spectrum is $S_{T_3 \otimes T_3} = (9,6,4,4,3,2,2,2,1)$, and the multi-kernel polar codes resulting from our design still outperform all other depicted designs. 

In Figure \ref{fig:plot_40} we show the performance of a code of length $N = 40$ and dimension $K = 20$. 
The transformation matrix is given by $T_{40}=T_2^{\otimes 3} \otimes T_5$, i.e., there is only one $T_5$ kernel at the rightmost of the Kronecker product. 
In this case, the BLER performance of the proposed construction is again better than the one of the other constructions, though the gain is smaller than in the previous two cases where $T_3$ is used. 

Finally, in Figure \ref{fig:plot_90} we show the performance of a code of length $N = 90$ and dimension $K = 45$. 
In this case, the transformation mixes three different kernels, and we define $T_{90}=T_2 \otimes T_3^{\otimes 2} \otimes T_5$. 
The BLER performance of the proposed construction, while is only able to match the performance of the shortened polar code. 
This shows that the proposed algorithm should be further optimized in the presence of multiple high size kernels. 

In conclusion, the proposed distance-based construction significantly outperforms state-of-the-art punctured and shortened polar codes for small block lengths, as well as the previously proposed reliability-based construction in \cite{mk_arxiv}. 
We expect this property to hold true for short code lengths, when the polarization effect has lower importance than the distance profile in the design of the codes. 
Moreover, we argue that the encoder and the decoder of the proposed multi-kernel polar codes have a lower complexity compared to the encoder and the decoder of the state-of-the-art punctured polar codes, due to the larger length of the mother polar code and the reliability calculations required for these constructions \cite{mk_arxiv}.

\section{Conclusions}
\label{sec:conclusions}
In this paper, we proposed a construction for multi-kernel polar codes, introduced in \cite{mk_arxiv}, based on the maximization of the minimum distance. 
While the original construction based on bit reliabilities is suitable for long codes, our new minimum-distance based construction provides significant performance gains for short codes, i.e., where the polarization effect is less important than distance properties. 
This gives fundamental insights for the design of multi-kernel polar codes of any length. 
We further introduced the minimum-distance spectrum of a transformation matrix, and we developed a greedy algorithm that finds the information set achieving this minimum distance. 
Simulations illustrate the competitive performance of our design for short-length codes. 


\bibliographystyle{IEEEbib}
\bibliography{polar_codes_bib}

\end{document}